\documentclass[twocolumn,prb]{revtex4}
\usepackage{graphics,graphicx}
\begin{document}
\title{The Realization of Artificial Kondo Lattices in Nanostructured Arrays}
\author{D.K.~Singh$^{1,2}$}
\author{M.T.~Tuominen$^{3}$}
\affiliation{$^{1}$National Institute of Standard and Technology, Gaithersburg, MD 20899, USA}
\affiliation{$^{2}$Department of Materials Science and Engineering, University of Maryland, College Park MD}
\affiliation{$^{3}$Department of Physics, University of Massachusetts, Amherst, MA 01003,USA}

\begin{abstract}
The interplay of magnetic energies in a Kondo lattice is the underlying physics of a heavy fermion system. Creating an artificial Kondo lattice system by localizing the moments in an ordered metallic array provides a prototype system to tune and study the energetic interplay while avoiding the complications introduced by random alloying of the material. In this article, we create a Kondo lattice system by fabricating a hexagonally ordered nanostructured array using niobium as the host metal and cobalt as the magnetic constituent. Electrical transport measurements and magnetoresistivity measurements of these artificial lattices show that the competing exchange coupling properties can be easily tuned by controlling the impurity percentage. These artificial Kondo lattice systems enable the exploration of an artificial superconductor which should lead to a deep understanding of the role of magnetism in unconventional superconductors. 
\end{abstract}

 \maketitle

The role of magnetism in unconventional superconductors--cuprates,\cite{Tranquida}  heavy fermions\cite{Fisk}  and the recently discovered Fe-based pnictides\cite{Cruz} --is a current area of intense debate for physicists and material scientists alike. Typically, the energy scale of magnetic excitations and the superconducting order parameter have a one-to-one correspondence.\cite{Cox,Mathur} In heavy fermion compounds, this magnetic energetics is reflected via the interplay of the localized Kondo interaction\cite{Kondo,Andrei} —the screening interaction between a single magnetic atom and surrounding conduction electrons—and the long-range magnetic order.\cite{Hewson}  Heavy fermion compounds of intermetallic rare earth materials have the natural tendency to create the Kondo interaction like phenomena between a chemically ordered lattice of local moments ($f$-band) and the compound’s conduction electrons ($s$ and $p$-bands).\cite{Hewson} 

The formation of Kondo lattice in rare earth intermetallic compounds (of cerium, ytterbium, uranium) lead to many surprising phenomena which includes the observation of unconventional superconductivity (for example UPd$_{2}$Al$_{3}$, CeRhIn$_{5}$)\cite{Sato,Saxena,Anderson,Park} and quantum critical phenomena (for example YbRh$_{2}$Si$_{2}$).\cite{Si,Clusters} Quantum critical phenomena (QCP) are accompanied by a second order quantum phase transition between a long range magnetic order regime, favored by Ruderman-Kittel-Kasuya-Yoshida (RKKY) interaction which results from an indirect exchange interaction between local moments mediated by conduction electrons, and the Kondo screening regime at T = 0 K.\cite{Doniach,Gegenwart} The competing energy scales involved in QCP and their scaling with temperature is argued to be directly related to the superconducting order parameter, gap energy $\Delta$, in some of these compounds and therefore the magnetic energy interplay is considered at the core of the occurrence of superconductivity.\cite{Fisk,Sato} The crystalline nature of a Kondo lattice provides an ideal setup for study. In this article we describe the creation of artificial magnetic crystals in nanostructured arrays. The hexagonal ordering of magnetic impurities (cobalt) in the host metal (thin niobium film) automatically creates a magnetic crystal. Controlling the percentage of magnetic impurities embedded in each site in the host metal acts as the tuning parameter to allow for the detailed investigation of the energetic interplay between Kondo and RKKY-type interactions (Fig. 1). Exchange coupling constants were estimated using electrical and magnetoresistance measurements of the artificially fabricated Kondo lattices. As shown in the schematics in Fig. 1, our technique leads to the implantation of Co particles inside the Nb film which is a different physical system from magnetic dot arrays on top of a substrate. In the latter case, conduction electrons do not interact with the magnetic impurity’s moment directly. 

\begin{figure}
\centering
\includegraphics[width=4.3cm]{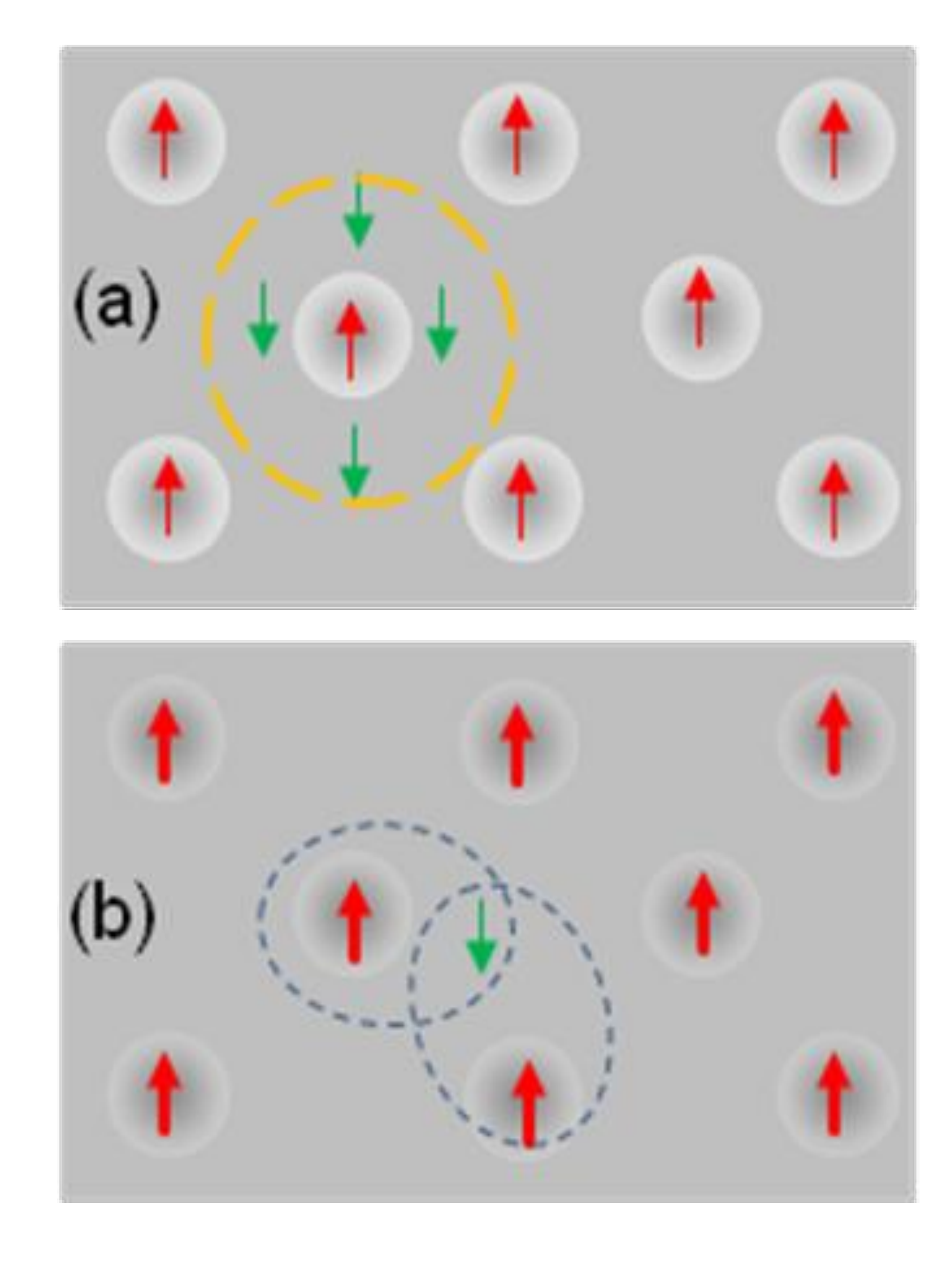} \vspace{-2mm}
\caption{(color online) Schematic description of artificial magnetic crystals. (a) As we see in this figure, localized magnetic ions (dark grey) embedded in a conducting film form a hexagonal lattice pattern and thus create an artificial magnetic crystal. If the amount of magnetic ions at each scattering site is small then each magnetic scattering site is totally screened by conduction electrons and therefore forms a dilute Kondo lattice system. (b) When the percentage of magnetic ions at each scattering site is increased with respect to the conducting film then indirect exchange interaction, mediated by conduction electrons, between localized magnetic ions plays a significant role. In figures (a) and (b), the percentages of magnetic ions at scattering sites are reflected by the (Red) arrow’s weight.
} \vspace{-4mm}
\end{figure}

The first step in the fabrication process of artificial Kondo lattices involves the development of hexagonal nanoporous copolymer templates from a self-assembled diblock copolymer film of thickness 36 nm and average pore diameter of 13 nm on top of a 10 nm thick conducting substrate (Nb). Lattice separation between pores was 28 nm. Details about the method of developing the nanoporous copolymer template can be found elsewhere.\cite{Albretch} The angular deposition of the desired material onto the walls of the nanopores was followed after this procedure. During the angular deposition process, the substrate (Nb film with nanoporous polymer template on top of a silicon wafer) was uniformly rotated about the axis perpendicular to its plane. Angular deposition depends on a critical deposition angle $\theta$$_{c}$ = tan$^{-1}$(D/h), where D and h are the diameter and height of a nanopore, respectively. In our experiment, a deposition angle of $\theta$ = 25 degree ($\geq$ $\theta$$_{c}$ = 20 degree) was chosen to avoid the deposition of Co at the bottom of the pores. The thickness of deposited Co material on the walls of nanopores was in the range of 4-5 nm (based on the calibrated quartz-crystal microbalance reading). This technique of angular deposition of Co onto the wall of the nanopores results in a multiply-connected geometry as the top of the film (polymer template) is always exposed to the evaporant, as shown in Fig. 2a and 2b. After material (cobalt) deposition, the calibrated parallel beam ion-beam etching technique was used to forcefully knock the desired amount of Co particles out of the nanopore’s walls and embed into the Nb substrate in an ultra-high vacuum chamber filled with argon gas at 10$^{-5}$ torr. For this purpose, the sample was tilted by 20$^{o}$ during the ion-beam etching and the ion beam was accelerated at 2000 eV. A schematic description of this process is shown in Fig. 2c-d. By controlling the etching rate and time, we were able to control the amount of Co particles embedded in a hexagonal order in Nb film. After etching, the sample was rinsed with toluene for 6 hrs to remove the remaining polymer template encapsulated with thin cobalt film, leaving only the Nb film with magnetic impurities. Fig. 2e-f show TEM (transmission electron microscope) images of artificial lattices resulting after 6 and 10 minutes of ion-beam etching. We clearly see in these figures that a certain percentage of Co particles are embedded in the host substrate and form a hexagonal matrix. For convenience, samples resulting after different etching times are named Nb-diblock (etching time = 0 min, equivalent to Fig. 2a), NC1 (etching time = 2 min), NC2 (etching time = 4 min), NC3 (etching time = 6 min) and NC4 (etching time = 10 min). Based on the calibrated ion-beam etching rate, it is roughly estimated that samples NC1, NC2, NC3 and NC4 have no more than 0.03$\%$, 0.06$\%$, 0.1$\%$ and 0.2 $\%$ of Co particles with respect to Nb, respectively. In terms of the number of Co particles per site, most dilute sample NC1 has about 25 Co particles per site. The uncertainty in the determination of the exact percentage of magnetic impurities embedded in the conducting substrate does not affect our conclusion. 

\begin{figure}
\centering
\includegraphics[width=10.0cm]{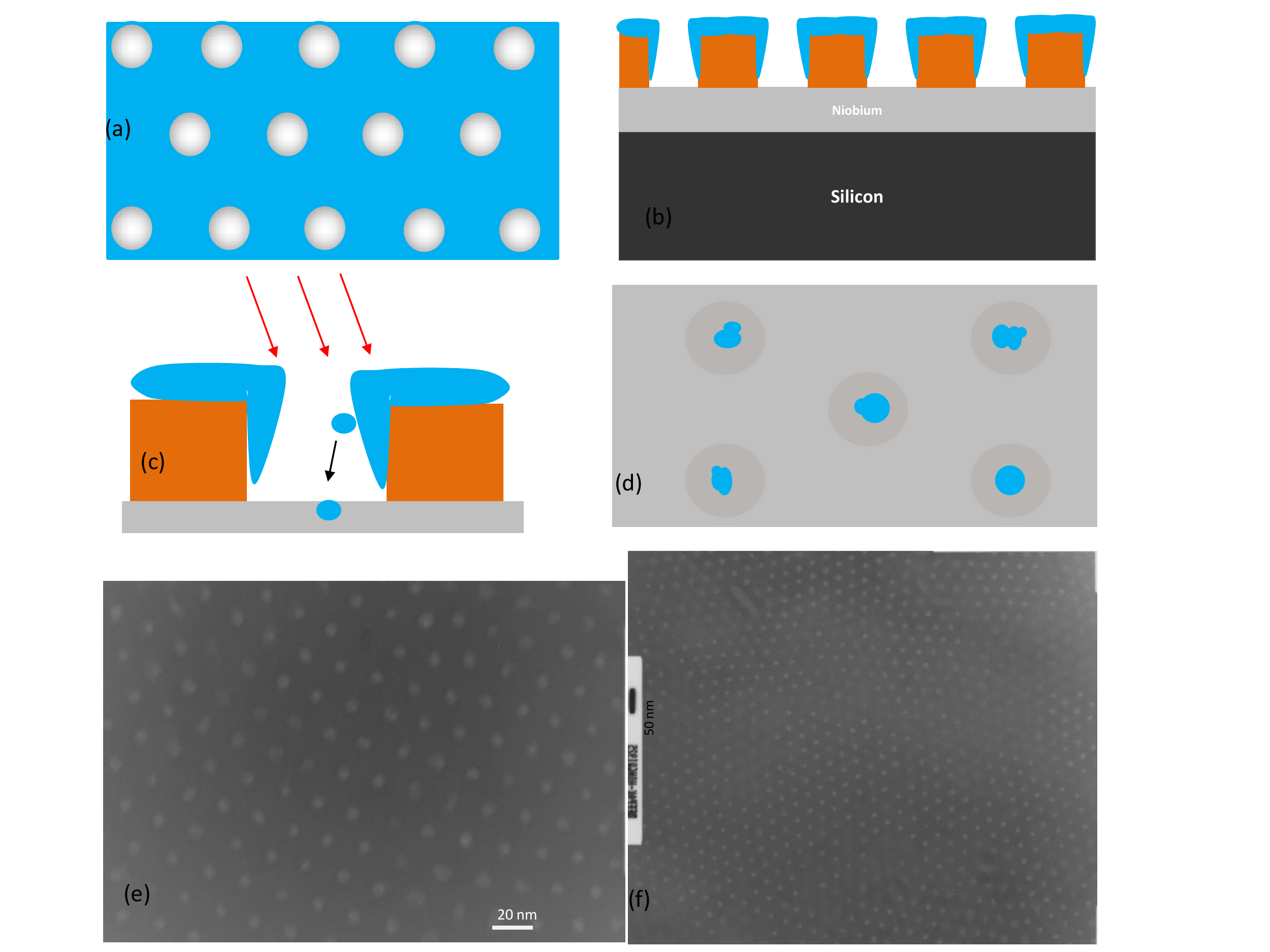} \vspace{-2mm}
\caption{(color online) Fabrication scheme and electron microscope images of artificial Kondo lattices. (a) $\&$ (b) Schematic views from top and side of multiply-connected hexagonally ordered diblock copolymer porous template encapsulated with cobalt film are shown in these figures. In this figure, orange color posts are diblock copolymer walls, white spaces are empty pores and blue capping is cobalt film, resulting from the angular deposition of cobalt on the wall of the pores. (c) Calibrated ion-beam etching technique is used to knock out desired amounts of cobalt particles from the encapsulating cobalt layer and forcefully implant them in the conducting substrate. (d) This schematic figure shows the top view of conducting film with cobalt particles embedded in hexagonal matrix after the ion-beam etching. (e) TEM image of sample NB3 (6 min. etching) is shown in this figure. We clearly see local hexagonal ordering of magnetic impurities in this figure. (f) TEM image of a large array of 10 min ion-beam etched film (NB4). 
} \vspace{-4mm}
\end{figure}

Electrical transport measurement data in the temperature range 4-300 K are shown below in Fig. 3a. Below 4 K, the substrate (Nb film with diblock polymer template only) becomes superconducting  and no useful information regarding the magnetic properties of the system can be extracted.\cite{Singh} Therefore we will focus on electrical measurements above 4 K only. Electrical resistivity data in Fig. 3a were normalized with respect to the value at T = 290 K to get a better comparison between different samples. Comparing the electrical transport data of all five samples, we observe different variations of the resistivity with temperature. The resistivity of the undoped Nb-diblock sample has been reported before and we will not discuss it here.\cite{Singh} Sample NC1 exhibits a deep Kondo-like resistance minimum around 70 K and saturates below 30 K as the measurement temperature is further reduced. The saturation of resistivity below T = 30 K suggests the onset of Kondo screening and formation of the Kondo lattice in this system.\cite{Das} This will be discussed in greater detail in the following paragraphs. In the case of NC2, the resistivity first increases, peaks around 200 K and then decreases before increasing again below $\simeq$ 40 K. The presence of the resistance minimum at T$\simeq$30 K suggests that the Kondo effect is still present, perhaps weakened, in this system. The broad maximum in resistance around 200 K is difficult to analyze due to possible contributions from many different effects such as thermal fluctuations and spin-diffusion. Nonetheless it is reasonable to argue that the resistance maximum at $\simeq$ 200 K corresponds to local spin correlations which exceed thermal disorder and begin to form magnetic impurity “clusters” via an indirect exchange interaction.\cite{Ford} No Kondo like resistance minimum is observed in sample NC3; instead the resistivity starts increasing as the measurement temperature is reduced untill T$\simeq$100 K. Below 100 K, the resistivity drops significantly before saturating below T$\simeq$30 K. Similar behavior of resistivity variation has been observed before in Kondo lattice systems and the onset temperature is often associated with the characteristic temperature indicating the transition to long range magnetic order mediated by the RKKY-type interaction.\cite{Walker,Stewart} Further increase in the percentage of magnetic impurities as a result of longer ion-beam etching leads to larger local moments. In that case, direct magnetic interaction between localized magnetic impurities starts playing a role. This sets up a new regime marked by the absence of Kondo like interaction and the presence of direct magnetic interaction between localized impurities, as observed in sample NC4 where an increase in resistivity at high temperature is followed by a gradual decrease as the temperature is lowered. To further elucidate the properties of magnetic crystals and learn about the interplay of Kondo interaction vs. RKKY interaction, the following paragraphs in this article will primarily focus on the resistive and magnetoresistive behavior of samples NC1, NC2 and NC3 only.

\begin{figure}
\centering
\includegraphics[width=9.0cm]{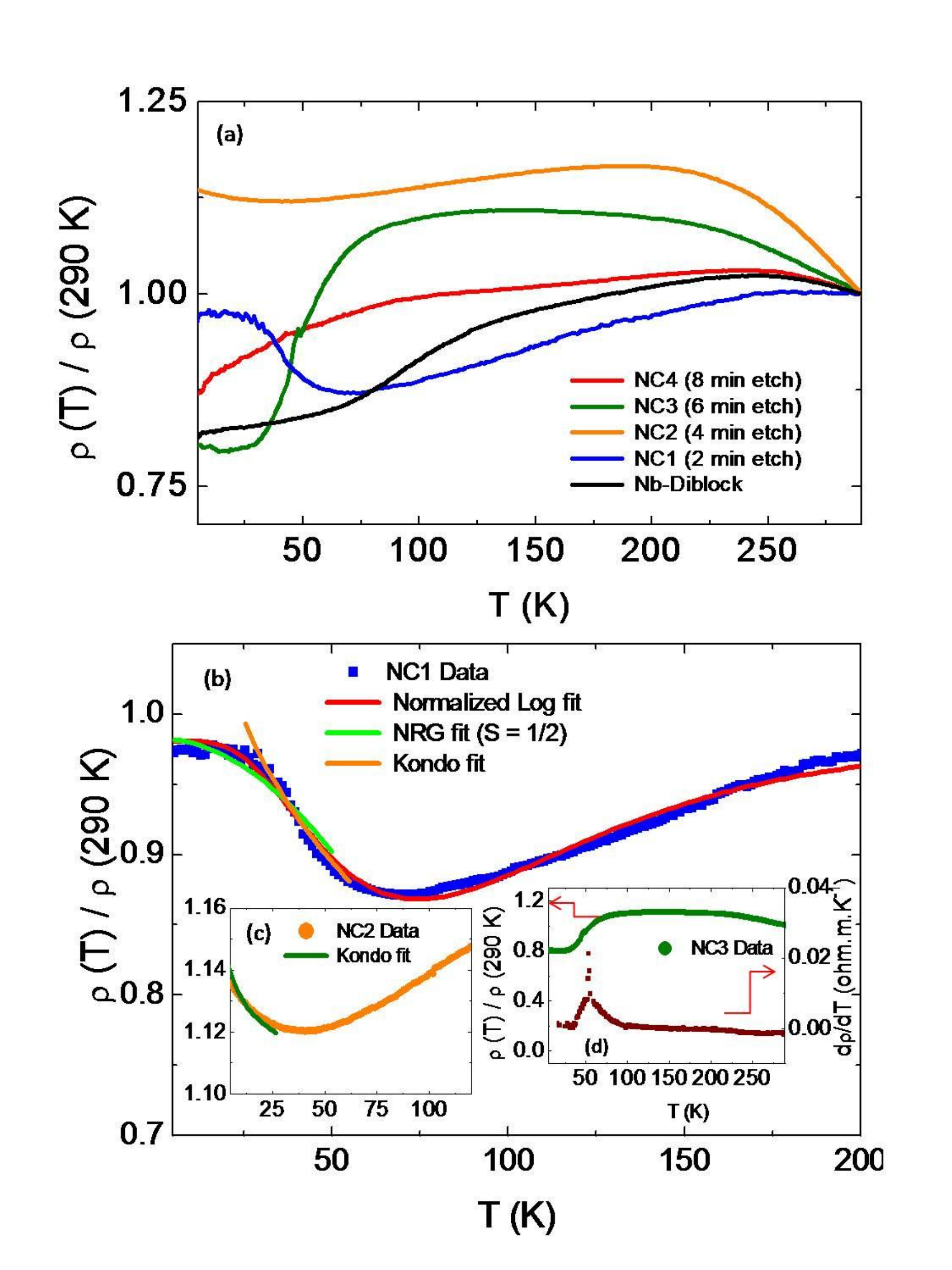} \vspace{-2mm}
\caption{(color online) Resistivity measurement analysis of magnetic crystals. (a) Normalized resistivity data of different magnetic crystals are shown in this figure along with the resistivity data of niobium film with diblock copolymer template only (reference sample). As the magnetic impurity percentage is increased at each scattering site, the physical behavior changes from Kondo screening (NC1) to RKKY-type interaction (NC3). (b) Resistivity data of sample NC1 is fitted with NRG soln. for spin-1/2 and Kondo expression. Obtained fitting parameters are reasonably consistent with each other (see text). Also shown in this figure is the log-normal behavior (Red curve) exhibited by NC1 over broad temperature range. (c) Low temperature data of sample NC2, fitted with Kondo expression, is shown in this figure. (d) Resistivity data and resistivity derivative data of sample NC3 are shown in this figure. A clear peak is observed in resistivity derivative data at T = 51 K while a large drop in resistivity starts occurring around T$\simeq$70 K.
} \vspace{-4mm}
\end{figure}

The Kondo interaction leads to the formation of composite fermions with spin quantum number S=1/2.\cite{Gegenwart} Theoretically it has been shown by Costi et al. that Wilson’s numerical renormalization group (NRG) method\cite{Wilson} can be effectively used to extract the characteristic Kondo temperature from transport measurements of dilute magnetically impure conducting systems.\cite{Costi,Mallet} The application of the NRG method to the Kondo model characterizes the physical properties of a Kondo system in which the impurity moment weakly couples to the conduction electrons. In Fig. 3b, we have fit the low temperature resistivity data of sample NC1 using the NRG solution for S = 1/2. As we see in this figure, the NRG solution for S = 1/2 gives a reasonable fit to the experimental data with T$_{K}$ = 56 K. To further verify the Kondo transition temperature, we have also fit the experimental data using the Kondo expression: $\rho$$\simeq$-ln(T/T$_{K}$) in the temperature range of T = 70 K to T = 30 K. The obtained fit of T$_{K}$, 56.7 K, is consistent with the NRG fitting value. These fits suggest that sample NC1 forms a Kondo lattice system with Kondo screening temperature of $\simeq$ 56 K. In the present analysis of experimental data, we ignore the weak-localization (WL) and electron-electron interaction (EEI) effects. In general, WL and EEI effects are more dominant in quasi one-dimensional systems where the dimensional length scale is smaller than the mean free path for conduction electrons, which is not the case in this work. Interestingly, the NC1 resistivity curve is also well fitted with log-normal behavior over a wide temperature range, as shown by the red curve in Fig. 3b. The obtained fitting value of T$_{K}$ = 59.2 K in this case is slightly larger than other fitting results for this system. This unusual log-normal behavior of resistivity as a function of temperature is observed for the first time in Kondo systems and suggests that Kondo screening and logarithmic normalization of resistivity are inter-related. 

We have also estimated the exchange interaction between the Co magnetic impurities and conduction electrons of Nb using the Kondo expression $K$$_{B}$$T$$_{K}$ $\simeq$ exp(-1/$J$$n$),\cite{Hewson} where $J$ is the absolute value of exchange interaction constant and n is the density of states of conduction electrons. For n=5.5$\times$10$^{28}$ $m$$^{-3}$,\cite{density} the estimated value of $J$ is found to be -9.7 meV. The estimated exchange constant for sample NC1 is of the same order as observed in other dilute Kondo lattice systems.\cite{Burdin} We have also fit the low temperature experimental data of NC2 using the logarithmic Kondo equation, as shown in Fig. 3c. The best fit is obtained for T$_{K}$ = 24 K which corresponds to $J$ = - 6.3 meV. It is noteworthy to mention that each hexagonal magnetic impurity location in NC2 has a higher percentage of Co as compared to NC1. That, possibly, prohibits the application of Kondo behavior to this system as the Kondo formalism is more applicable to dilute impurity cases. Increasing the concentration of magnetic impurities also leads to the onset of indirect exchange interaction between impurities.

Sample NC3 does not exhibit Kondo-type minima in resistivity measurements; rather a large drop in resistivity is observed around 75 K. We have also plotted d$\rho$/dT vs. T in Fig. 3d. A clear peak is observed in differential resistivity data at T = 51 K which quantifies the point of inflection of the resistivity curve. Similar behavior in resistivity measurements has been observed before in concentrated Kondo lattice systems and were associated with the RKKY-type indirect exchange interaction between magnetic impurities mediated by conduction electrons.\cite{Schilling} Since RKKY-type interactions between magnetic ions leads to long range magnetic order, many researchers argue that the resistivity maxima represent the magnetic transition temperature, T$_{M}$.\cite{Walker} Since we are interested in estimating the exchange interaction strength of RKKY-type interaction for sample NC3, we chose the point of inflection of the resistivity curve, Fig. 3d, as the characteristic temperature, T$_{RKKY}$ = 51 K, below which conduction electron mediated RKKY-type interaction plays a significant role. RKKY exchange coupling strength, $J$$_{RKKY}$, is estimated using the expression $K$$_{B}$$T$$_{RKKY}$ $\simeq$ $J$$^{2}$$n$.\cite{Yang} The estimated magnitude of the exchange coupling constant $J$$_{RKKY}$ is approximately 2.1 meV. Thus we see that the exchange coupling strength changes as the amount of magnetic impurities is varied from dilute (2 min. etching) to concentrated (6 min etching) in the hexagonally localized matrix. 

To further understand the magnetic properties of Kondo lattices, we have performed magnetoresistance measurements for both in-plane and perpendicular application of magnetic fields with respect to the plane of the substrate. In Fig. 4a-c, the magnetoresistance data of NC1, NC2 and NC3 are plotted as a function of temperature. Magnetoresistance (MR) is defined as MR = [$\rho$(H) - $\rho$(0)]/$\rho$(0). As we can see in Fig. 4a, the application of an external magnetic field up to 2 Tesla on sample NC1 does not change the MR substantially. Also, no observable differences between MR data for in-plane and perpendicular application of magnetic fields are found. This simply indicates that the magnetic impurity spins are completely screened by surrounding conduction electrons and hence remain unaffected by the application of an external magnetic field. On the other hand, the magnetoresistance data of sample NC2 (Fig. 4b) exhibit different responses for magnetic field applied in-plane and perpendicular to the substrate. MR value at a particular temperature is larger at higher magnetic field and overall, magnetoresistance curves for perpendicular field application lie higher compared to MR curves for in-plane field application. The perpendicular direction to the substrate, in this case, evolves as the easy axis.

\begin{figure}
\centering
\includegraphics[width=9.0cm]{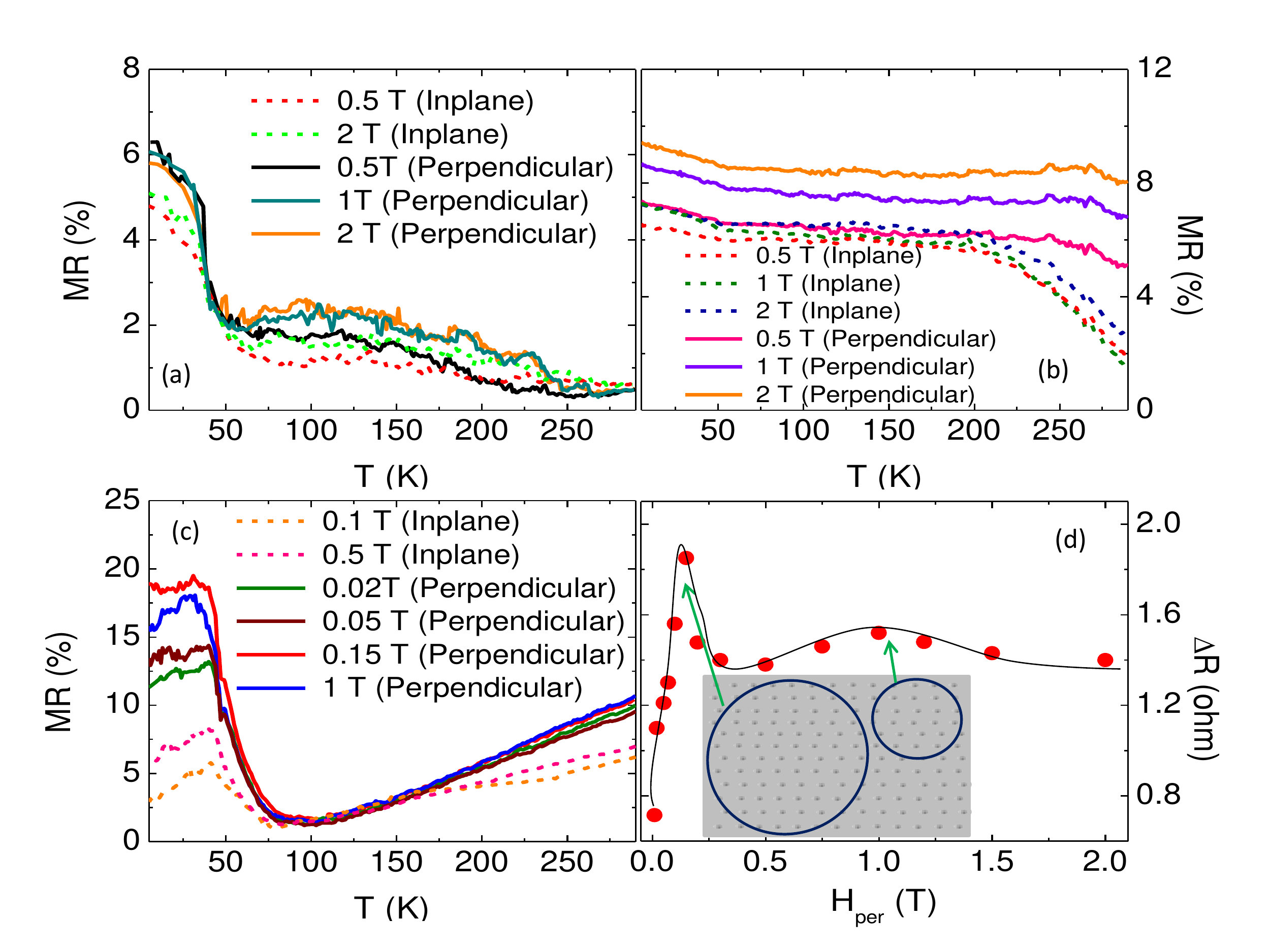} \vspace{-2mm}
\caption{(color online) Magnetoresistance (MR) measurements of Kondo lattices. (a) Magnetoresistance data of sample NC1 for both in-plane and perpendicular applications of external magnetic field are plotted in this figure. No clear difference is observed between measured data at different applied fields, indicating complete Kondo screening of localized moments. (b) In the case of sample NC2, MR data for perpendicular application of magnetic fields lie higher compared to in-plane data of same field values. It suggests that Kondo screening is weakened and the direction perpendicular to the substrate develops as the easy axis for this system. (c) In this figure, we have plotted the MR data of sample NC3. Significantly higher MR percentages are observed at low temperature in this case. We also observe a deep minimum in MR indicating the onset of coherence and heavy fermion state (see text). (d) We have plotted the change in resistance, $\Delta$$R$, as a function of perpendicular magnetic field at T = 7 K in this figure. Solid line is a guide to the eye. Observed peaks in the $\Delta$$R$ data are attributed to the cyclotron radius of A-B rings circling different number of scattering centers (as shown in the inset).
} \vspace{-4mm}
\end{figure}

In the MR measurements of NC3, two interesting behaviors are observed: the formation of MR minima around $\simeq$ 80 K which suggests the onset of coherence and heavy fermion state\cite{Groten} and the fluctuation of magnetoresistance as a function of perpendicular magnetic field. MR value for perpendicular application of field in this case is as high as $\simeq$20 $\%$ at low temperature. In Fig. 4d, we have plotted the change in resistance $\Delta$$R$ = $R(H)$-$R(0)$ as a function of perpendicular applied field at $T$ = 7 K. We clearly see the oscillatory nature of $\Delta$$R$ as a function of field in this figure. In general, localized magnetic impurities act as the scattering centers for conduction electrons which lead to interference effects between conduction electrons when the phase coherence of electrons exceeds the inter-impurity distance.\cite{Thornton} If the magnetic impurities are arranged in a regular fashion then conduction electrons form Aharonov-Bohm (A-B) rings encircling a discrete number of impurities.\cite{Weiss} Radius ($r$$_{AB}$) of A-B rings are given by $\Delta$$B$$_{\perp}$ $\simeq$ h/($\pi$$e$$r$$_{AB}$$^{2}$). In Figure 4d, we clearly see a strong peak in magnetoresistance data at H$\simeq$ 0.15 T and a relatively weaker peak at H$\simeq$1 T. Using the above expression for A-B ring’s radius, these magnetic field values correspond to A-B rings of 300 nm and 120 nm radiuses respectively. These cyclotron radiuses correspond to electrons encircling 64 and 24 magnetic scattering centers respectively, as shown in the schematic description in the inset of Figure 4d. Recently Siegert et al. identified similar experimental observation in a semiconductor heterostructure with RKKY-type interaction by correlating the oscillatory behavior of indirect exchange constant, $J$, with the A-B radius r$_{AB}$ via range function over a large distance ($\simeq$500 nm).\cite{Siegert}

The energetic interplay between the small exchange energy Kondo screening regime and the large exchange energy RKKY-type regime can also be easily tuned by changing the lattice separation between magnetic impurities. The nature of RKKY indirect exchange interaction varies between ferromagnetic and antiferromagnetic interactions with distance.\cite{Skomski} Thus the combination of the change in magnetic impurities percentage with the variation in lattice separation can be used to tune both the strength and nature of energetic interplay. Experiments of artificial Kondo lattices were also repeated with aluminum as the host metal and similar resistivity patterns were observed. However, the estimated exchange constants in different regimes (Kondo screening and RKKY-type) were found to be relatively larger. As the thickness of the metallic host was increased to 25 nm, a sharp downturn in the electrical resistance data was observed below the resistance maxima. This suggests that the energetic interplay can be further explored by choosing different combinations of host metal and magnetic impurities. It will be interesting to explore the scaling behavior in these artificial Kondo lattice systems, in comparison to those recently reported by Yi-Feng et al. in inter-metallic rare earth compounds.\cite{Yang} The scaling of Kondo lattices involve a characteristic temperature, T$^{*}$, defined by the intersite coupling energy.\cite{Nakatsuji} Precise estimation of T$^{*}$ in these magnetic crystals would require magnetic susceptibility and heat capacity measurements.

We now discuss the important implication of artificial Kondo lattice creation. Many Kondo lattice compounds exhibit the competing dual phenomena of unconventional superconductivity and magnetic energy interplay.\cite{Sato,Saxena,Park} However, the observation of superconductivity is not universal in all Kondo lattice systems. This prohibits any strong theoretical formulation combining magnetic energy interplay and superconductivity. Here we propose a reverse approach using artificial Kondo lattice systems. We have shown that the energetic interplay of the competing magnetic interactions occurs in the range of -9.7 meV to 2.1 meV in the presented artificial Kondo lattices, similar to that in the bulk Kondo lattice systems exhibiting unconventional superconductivity. If the magnetic energy interplay and superconductivity are directly related and the magnetic excitations provide a mediating mechanism for Cooper pair formation, essential for superconductivity, then it would be possible to create an artificial crystal with superconducting properties. Additional study of the artificial magnetic crystals using different metallic host with higher electronic density of states and different magnetic constituents are necessary for this purpose. Discovery of superconductivity in an artificial Kondo lattice system will not only firmly establish the role of magnetism in unconventional superconductors but will also provide an important tool to manipulate the physical properties of the materials, which will be crucial for the future development of higher temperature superconductors.

We thank J. W. Lynn, J. Gardener, S. Chang and S. Chi for helpful discussions and comments. This work was supported by NSF Grants DMR-0531171, DMR-0306951 and MRSEC.

\bibliography{AKL}
\end{document}